\documentclass[preprint,pra,showpacs,showkeys]{revtex4}
\usepackage{graphicx,amsmath,amsfonts,amssymb}
\usepackage{times}

\begin{document}

\title{Entanglement dynamics of two-qubit system in
different types of noisy channels \footnote{Supported by  the
Natural Science Foundation of Hubei Province, China under Grant No
2006ABA055, and the Postgraduate Programme of Hubei Normal
University under Grant No 2007D20.}}

\author{C. J. Shan\footnote{ E-mail: scj1122@163.com}}
\author{J. B. Liu}
\author{W. W. Cheng}
\author{T. K. Liu\footnote{Corresponding author. E-mail:
tkliuhs@163.com}}
\author{Y. X. Huang}
\author{H. Li}
\affiliation{College of Physics and Electronic Science, Hubei Normal
University, Huangshi 435002, China}
\date{\today}

\begin{abstract}

In this paper, we study entanglement dynamics of a two-qubit
extended Werner-like state locally interacting with independent
noisy channels, i.e., amplitude damping, phase damping and
depolarizing channels. We show that the purity of initial entangled
state has direct impacts on the entanglement robustness in each
noisy channel. That is, if the initial entangled state is prepared
in mixed instead of pure form, the state may exhibit entanglement
sudden death (ESD) and/or be decreased for the critical probability
at which the entanglement disappear.
\end{abstract}



\maketitle

\section{Introduction}

In the last two decades, entanglement attracts much attention due to
the powerful applications in quantum information process and quantum
computing.$^{[1,2]}$ In order to realize quantum information
process, great effort has been devoted to studying and
characterizing the entanglement in cavity QED $^{[3-5]}$ and spin
systems$^{[6,7]}$ schemes. Therefore, it is of increasingly
importance to understand entanglement behaviors of quantum system in
realistic situations, where the system unavoidably loses its
coherence due to interactions with the environment. In this context,
a peculiar dynamical feature of entanglement was discovered that the
entanglement can vanish completely in a finite time, in striking
contrast with decoherence of its individual constituent that decays
only asymptotically. Such a surprising phenomenon was termed
entanglement sudden death (ESD)$^{[8]}$. Because of its intrinsic
and practical interests, ESD has attracted much attention in
theory$^{[9-14]}$ and confirmed experimentally$^{[15]}$ for the case
of two qubits. The ESD phenomenon illustrates the fact that the
global behavior of an entangled system may be markedly different
from the individual and local behavior of its constituents.$^{[14]}$

From a practical point of view, an entangled state undergoing ESD
would put a limitation on the time of its application in practice
since it is less robust than one without ESD. Hence, to find various
conditions under which the ESD occurs seems to be very necessary. It
has shown that ESD is sensitive to the type of initial entanglement,
i.e., depending on the type of initial state, the entanglement may
or may not exhibit ESD.$^{[10]}$ Besides the initial condition of an
entangled state, the environment is another decided factor
responsible for the dynamical behaviors of entanglement. In Ref.
[14], the authors studied entanglement dynamics in three types of
noisy channels. It was shown that for the same entangled states,
their entanglement may exhibit completely different behaviors
involving the appearance of ESD in different types of environments.
However, they only considered the pure case of initial entangled
states. In this paper, we shall study the entanglement dynamics of
an entangled state subject to various noisy channels by paying more
addition to the initial condition of the state. We show that the
purity of initial entangled state is closely related to the
entanglement robustness in each noisy channel. That is, if the
initial entangled state is prepared in mixed instead of pure form,
the state may exhibit ESD and/or be decreased for the critical
probability at which the entanglement disappear.

To quantify the entanglement of a two-qubit system we adopt
Wootters' concurrence$^{[16]}$. The concurrence $C(\rho _{AB})$ for
the density matrix $\rho _{AB}$ of a two-qubit system $AB$ is
defined as
\begin{equation}
C(\rho _{AB})=\max \{0,\sqrt{\lambda _{1}}-\sqrt{\lambda
_{2}}-\sqrt{\lambda _{3}}-\sqrt{\lambda _{4}}\},  \label{C}
\end{equation}
where $\lambda _{i}$ are the eigenvalues of the matrix $\zeta =\rho
(\sigma _{y}^{A}\otimes \sigma _{y}^{B})\rho ^{*}(\sigma
_{y}^{A}\otimes \sigma _{y}^{B})$ arranged in decreasing order. Here
$\sigma _{y}^{A(B)}$ are the $y $-Pauli matrix acting on qubit $A$
$(B)$ and $\rho ^{*}$ is the complex conjugation of $\rho $ in the
standard (computational) basis. For the separate state $C=0$ whereas
$C=1$ for maximally entangled state.

For the initial state of qubit-pair $AB$, instead of Bell-like and
Werner states$^{[17]}$, we shall consider the following extended
Werner-like state
\begin{equation}
\rho _{AB}(0)=r\left| \Phi \right\rangle _{ABAB}\left\langle \Phi
\right| +\frac{1-r}{4}I_{AB}, \label{Wern}
\end{equation}
with $r$ the purity of the initial state of qubits $AB$, $I_{AB}$
the $4\times 4$ identity matrix and
\begin{equation}
\left| \Phi \right\rangle _{AB} =\left( \sin\theta\left|
00\right\rangle +\cos\theta\left| 11\right\rangle \right) _{AB},
\label{Bell}
\end{equation}
the Bell-like state. Obviously, the state in Eq. (\ref{Wern})
reduces to the standard Werner state when $\theta =\pi /4$ and to
Bell-like pure state when $r=1$. By dealing with the above extended
Werner-like state, we are able to study the effect of mixedness of
the initial entangled state. Both the Bell-like state and Werner
state, and so the extended Werner-like state, belong to the so-called $X$%
-class state$^{[12]}$ whose density matrix is of the form
\begin{equation}
\rho _{AB}=\left(
\begin{array}{cccc}
x & 0 & 0 & v \\
0 & y & u & 0 \\
0 & u^{*} & z & 0 \\
v^{*} & 0 & 0 & w
\end{array}
\right) ,  \label{X}
\end{equation}
with $x,y,z,w$ real positive and $u,v$ complex quantities. The $X$%
-class states have the property that the corresponding two-qubit
density matrix preserves the $X$-form during the system evolution.
For the $X$-state (\ref{X}), the concurrence can be derived as
\begin{equation}
C(\rho _{AB})=2\max \{0,|u|-\sqrt{xw},|v|-\sqrt{yz}\}.  \label{CX}
\end{equation}

We consider three paradigmatic types of noisy channels, i.e.,
amplitude damping channel, phase damping channel and depolarization
channel. Each one of the two qubits $A$ and $B$ individually coupled
to its own noisy environment, implying no any interaction, direct or
indirect, between them. The dynamics of each qubit is governed by a
master equation that gives rise to a completely positive
trace-preserving map $\mathcal{E}_{i}$ (with $i=A,B$) describing the
evolution as $\rho_{i}=\mathcal{E}_{i}\rho_{i}(0)$, where
$\rho_{i}(0)$ and $\rho_{i}$ are, respectively, the initial and
evolved reduced states of the $i$-th subsystem. In the following, we
shall consider the entanglement dynamics of the state (\ref{Wern})
in the three noisy channels, respectively.

\section{ Amplitude damping channel}

The first noisy channel we consider is the amplitude damping (AD)
channel which can be represented via the Kraus representation
as$^{[1,14]}$
\begin{equation}\label{kraus}
\mathcal{E}_{i}^{AD}\rho_{i}=E_{0}\rho_{i}E_{0}^{\dag}+E_{1}\rho_{i}E_{1}^{\dag},
\end{equation}
with $E_{0}=\left|0\right\rangle\left\langle0\right|+\sqrt{1-p}
\left|1\right\rangle\left\langle1\right|$ and $E_{1}=\sqrt{p}
\left|0\right\rangle\left\langle1\right|$ being its Kraus operators.
$p\equiv p(t)\equiv1-e^{-\frac{1}{2}\gamma t}$ is the probability of
the qubit exchanging a quantum with the bath at time $t$, and
$\gamma$ is the zero-temperature dissipation rate. After
time-evolving, the initial state of qubits $A,B$ in Eq. (\ref{Wern})
evolves into $\rho _{AB}(p)$ which has the $X$-form (\ref{X}) with
the matrix elements are given as
\begin{eqnarray}
x &=&\frac{1-r}{4}(1+p)^2+r(\cos ^{2}(\theta )p^2+
\sin ^{2}(\theta )), \nonumber \\
y &=&z=\frac{1-r}{4}(1-p^2)+r\cos ^{2}(\theta )p(1-p), \nonumber \\
w &=&(\frac{1-r}{4}+r\cos ^{2}(\theta ))(1-p)^2,  \nonumber \\
v &=&r\sin\theta\cos\theta(1-p),  \nonumber \\
u &=&0.
\end{eqnarray}
By virtue of Eq. (\ref{CX}), we can get the concurrence $C(\rho
_{AB}(p))$ of $\rho _{AB}(p)$ as
\begin{equation}
C(\rho _{AB}(p))=2\max \{0,|v|-\sqrt{yz}\}.  \label{Cab}
\end{equation}

If the initial state of qubits $A,B$ is pure, i.e., $r=1$, from the
relation $|v|-\sqrt{yz}=0$ we can get the critical probability
$p_{c}$ at which the entanglement disappear as
$p_{c}=|\frac{\sin\theta}{\cos\theta}|$. For $\theta<\pi/4$, $p_{c}$
is always smaller than 1, meaning that the entanglement disappears
before the steady state is asymptotically reached.$^{[14]}$ Thus
$\theta<\pi/4$ is the condition for the occurrence of ESD when the
initial state of $A,B$ is pure, i.e. $r=1$. The relations between
$C(\rho _{AB}(p))$ and $p$, $\theta$ are plotted in FIG. 1 for
$r=1$. However, we shall pay more attention to the case of $r<1$,
i.e., the initial entangled state is prepared mixed form. In this
case, the critical probability
$p_{c}=\frac{4r|\sin\theta\cos\theta|+r-1}{4r|\cos^2\theta|-r+1}$,
which is related to both the degree and purity of initial
entanglement in terms of $\theta$ and $r$. The condition of
appearance of ESD is derived as
$|\sin\theta\cos\theta|-\cos^2\theta<\frac{1}{2}(\frac{1}{r}-1)$,
from which we can see that the range of $\theta$ for appearing ESD
depends on the value of $r$. As an example, for $r=0.7$ we plot in
FIG. 2 the concurrence $C(\rho _{AB}(p))$ as functions of $p$ and
$\theta$, where we can see that ESD occurs for all the possible
values of $\theta$, in striking contrast with the case of $r=1$ in
FIG. 1. Hence, ESD phenomenon is closely related to the purity of
initial entangled state.

\begin{figure}[tbp]
\centerline{\scalebox{0.9}{\includegraphics{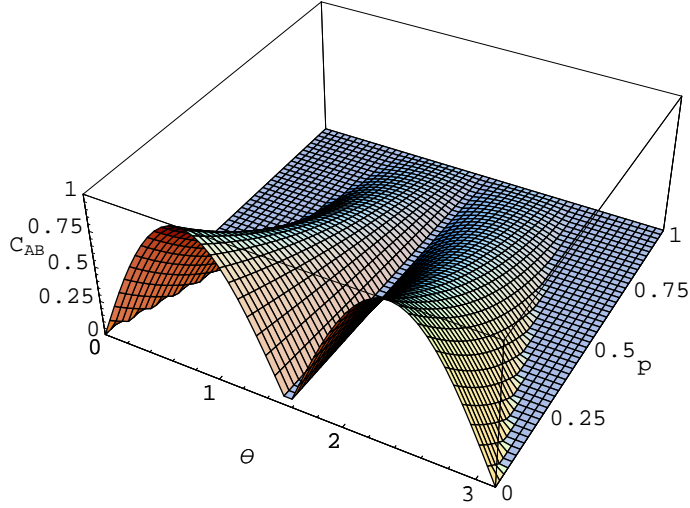}}}
\caption{$C_{AB}\equiv C(\rho _{AB}(p))$ as functions of $\theta$
and $p$ for the case of $r=1$. The ESD appears for
$\theta<\pi/4+k\pi$.} \label{fig1}
\end{figure}

\begin{figure}[tbp]
\centerline{\scalebox{0.9}{\includegraphics{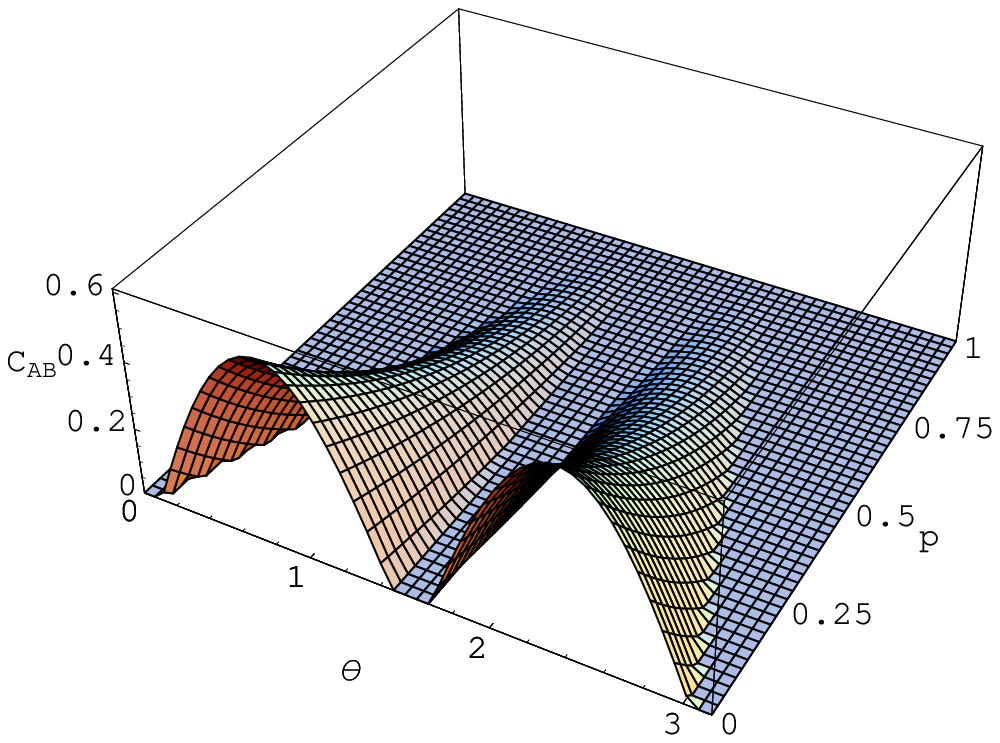}}}
\caption{$C_{AB}\equiv C(\rho _{AB}(p))$ as functions of $\theta$
and $p$ for the case of $r=0.7$. The ESD appears for all the
possible values of $\theta$.} \label{fig2}
\end{figure}

\section{ Phase damping channel}

In this section, we consider the phase damping channel (PD) in which
there is loss of quantum information with probability $p$, but
without any energy exchange. It is defined as$^{[14]}$
\begin{equation}\label{PDC}
\mathcal{E}_{i}^{PD}\rho_{i}=(1-p)\rho_{i}+p(|0\rangle\langle0|\rho_{i}|0\rangle\langle0|
+|1\rangle\langle1|\rho_{i}|1\rangle\langle1|).
\end{equation}
After time-evolving, the diagonal terms of initial state in Eq.
(\ref{Wern}) of qubits $A,B$ remains the same, whereas the
off-diagonal ones are multiplied by $(1-p)^2.$ By virtue of Eq.
(\ref{CX}), we can get the concurrence $C(\rho _{AB}(p))$ of $\rho
_{AB}(p)$ as
\begin{equation}
C(\rho _{AB}(p))=2\max \{0,|v|-\sqrt{yz}\},
\end{equation}
with $y=z=\frac{1-r}{4}$ and $v=r\sin\theta\cos\theta(1-p)^2$.
Obviously, if the initial state of qubits $AB$ is pure, i.e., $r=1$,
we have $y=z=0$, thus $C(\rho
_{AB}(p))=2\sin\theta\cos\theta(1-p)^2\geq0$ with equality hold for
$p=1$ implying non-existence of ESD for all the possible values of
$\theta$. However, if the initial state of qubits $AB$ is mixed the
situation will be different. From the relation $|v|-\sqrt{yz}=0$ we
can get the critical probability $p_{c}$ as
$p_{c}=1-\sqrt{\frac{1-r}{4r|\sin\theta\cos\theta|}}$. For $r<1$,
$p_{c}$ is always smaller than 1, meaning that in the PD channel the
entangled state (\ref{Wern}) always suffers ESD if the state is
prepared in the mixed form. To make a comparison, in FIG. 3 and FIG.
4, we plot $C(\rho _{AB}(p))$ as functions of $\theta$ and $p$ for
$r=1$ and $r=0.7$, respectively. Hence, the close relation between
ESD and the purity of initial entangled state can also be obtained
in the PD channel.

\begin{figure}[tbp]
\centerline{\scalebox{0.9}{\includegraphics{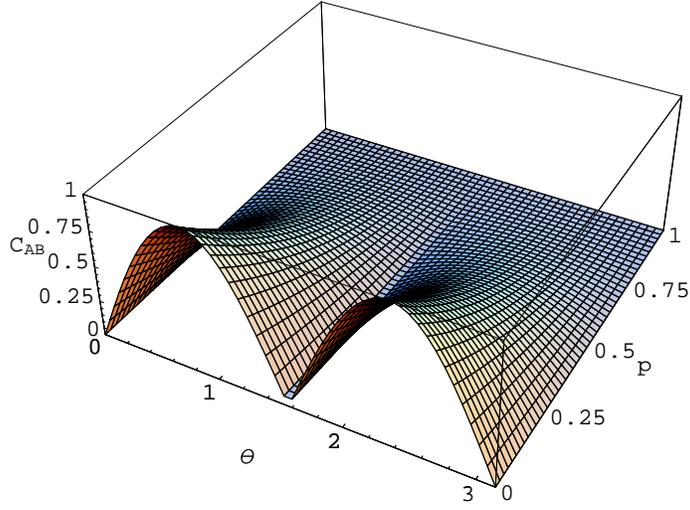}}}
\caption{$C_{AB}\equiv C(\rho _{AB}(p))$ as functions of $\theta$
and $p$ for the case of $r=1$. The ESD does not appear for all the
possible values of $\theta$.} \label{fig3}
\end{figure}

\begin{figure}[tbp]
\centerline{\scalebox{0.9}{\includegraphics{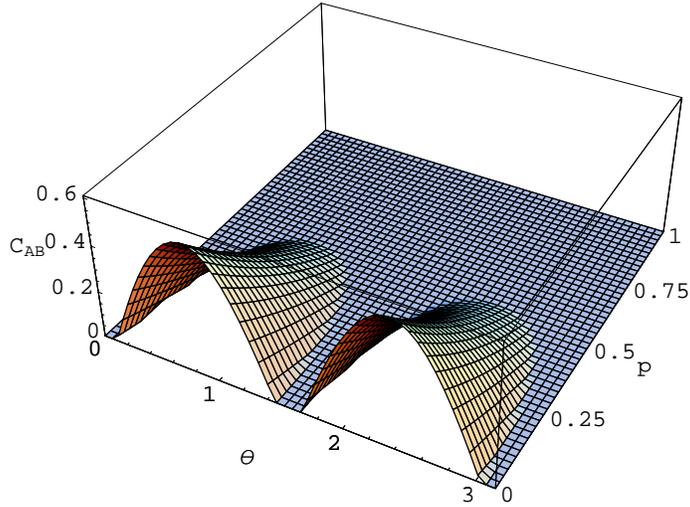}}}
\caption{$C_{AB}\equiv C(\rho _{AB}(p))$ as functions of $\theta$
and $p$ for the case of $r=0.7$. The ESD instead appears for all the
possible values of $\theta$.} \label{fig4}
\end{figure}

\section{ Depolarizing channel}

Next, we consider entanglement dynamics in the depolarizing (D)
channel. The D channel represents the situation where the $i$-th
qubit remains untouched with probability $1-p$, or is
depolarized-meaning that its state is taken to the maximally mixed
state-with probability $p$. It can be expressed as$^{[14]}$
\begin{equation}\label{DC}
\mathcal{E}_{i}^{D}\rho_{i}=(1-p)\rho_{i}+p\frac{\mathbf{I}}{2},
\end{equation}
with $\mathbf{I}$ is the identity operator.

After time-evolving, the density matrix (\ref{Wern}) evolves into
$\rho_{AB}(p)$ which remains the $X$ form (\ref{X}) with the matrix
elements are given as
\begin{eqnarray}
\nonumber x &=& (1-\frac{p}{2})^2(\frac{1-r}{4}+r\sin^2\theta)
+p(1-\frac{p}{2})\frac{1-r}{4}+(\frac{p}{2})^2(\frac{1-r}{4}+r\cos^2\theta), \\
\nonumber y &=&
z=\frac{1-r}{4}(1-p+\frac{p^2}{2})+\frac{p}{4}(1-\frac{p}{2})(1+r),\\
\nonumber w &=& (1-\frac{p}{2})^2(\frac{1-r}{4}+r\cos^2\theta)
+p(1-\frac{p}{2})\frac{1-r}{4}+(\frac{p}{2})^2(\frac{1-r}{4}+r\sin^2\theta), \\
\nonumber v &=&r\sin\theta\cos\theta(1-p)^2,   \\
u &=&0.
\end{eqnarray}
By virtue of Eq. (\ref{CX}), we can get the concurrence $C(\rho
_{AB}(p))$ of $\rho _{AB}(p)$ as
\begin{equation}
C(\rho _{AB}(p))=2\max \{0,|v|-\sqrt{yz}\}.
\end{equation}
For $r=1$ the state (\ref{Wern}) always undergoes ESD for all the
possible values of $\theta$ in the time evolution process as shown
in FIG. 5, where we plot $C(\rho _{AB}(p))$ as functions of $p$ and
$\theta$. It is known that ESD is a representation for the fragility
of entanglement. If we take the mixedness of initial entanglement
into account, the robustness of entanglement will be reduced
further. In FIG. 6 we plot $C(\rho _{AB}(p))$ as function of $p$ for
different $r$, where we can see that the critical probability will
be decreased with $r$.

\begin{figure}[tbp]
\centerline{\scalebox{0.9}{\includegraphics{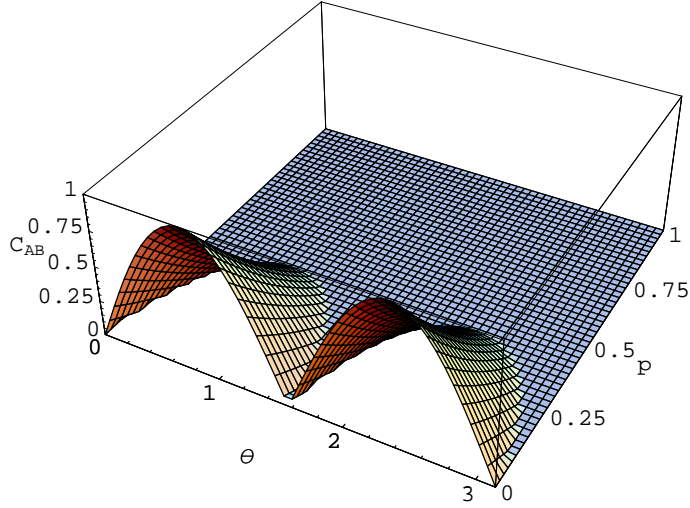}}}
\caption{$C_{AB}\equiv C(\rho _{AB}(p))$ as functions of $\theta$
and $p$ for the case of $r=1$. The ESD happens for all the possible
values of $\theta$.} \label{fig5}
\end{figure}

\begin{figure}[tbp]
\centerline{\scalebox{0.9}{\includegraphics{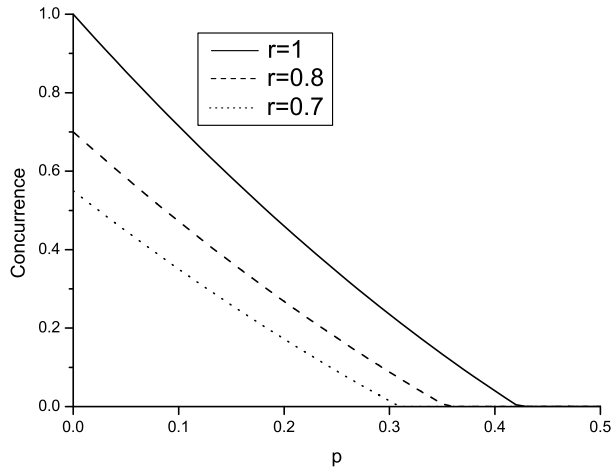}}}
\caption{$C(\rho _{AB}(p))$ as function of $p$ for $\theta=\pi/4$
for various $r$. The critical probability decreased with $r$.}
\label{fig6}
\end{figure}

\section{ Conclusions}

In conclusion, we have studied the entanglement evolution of a
two-qubit entangled state subject to independent environments, i.e.,
the amplitude damping, phasing damping and depolarizing channels.
The initial entangled state is prepared in an extended Werner-like
form (\ref{Wern}), thus we can investigate the relations between the
purity of the entangled state and its dynamical behaviors. We find
that the dynamical behaviors of the two-qubit entanglement are
closely related to its purity as well as the types of noisy
channels. The mixedness has direct impacts on the robustness of
entanglement in the sense that it can result in ESD and/or decrease
the critical probability. The understanding of entanglement
behaviors in realistic system is a precondition for its application
in practice, thus studying the conditions which may influence the
entanglement dynamics in various situations prove very important and
necessary.\\
We thank Z-X Man for his reading of the manuscript.

\end{document}